\documentclass[rnote]{aa}  

\usepackage{graphicx}
\usepackage{txfonts}
\usepackage{natbib}
\begin{document}
   \title{Microturbulent velocity from stellar spectra: \\
   a comparison between different approaches}
    \authorrunning{A. Mucciarelli}
  
   \titlerunning{Microturbulent velocity in stellar spectra}

   \author{A. Mucciarelli
          \inst{1}
          }

   \institute{Dipartimento di Astronomia, Universit\'a degli Studi 
              di Bologna, via Ranzani 1, 40126, Bologna, Italy\\
              \email{alessio.mucciarelli2@unibo.it}
             }

   \date{Received ....; accepted .....}

% \abstract{}{}{}{}{} 
% 5 {} token are mandatory
 
  \abstract
   {The classical method to infer microturbulent velocity in stellar 
   spectra requires that the abundances of the iron lines are not 
   correlated with the observed equivalent widths. 
   An alternative method, requiring the use of the expected line strength, 
   is often used to by-pass the risk of spurious slopes due to the 
   correlation between the errors in abundance and equivalent width.
  }
   {To compare the two methods and identify pros and cons and applicability to the 
   typical practical cases.}
   {I performed a test 
   with a grid of synthetic spectra, including instrumental broadening 
   and Poissonian noise. For all these spectra, microturbulent velocity 
   has been derived by using the two approaches and compared with the 
   original value with which the synthetic spectra have been generated.}
   {The two methods provide similar results for spectra with SNR$\ge$70, 
   while for lower SNR both approaches underestimate the true microturbulent 
   velocity, depending of the SNR and the possible selection of the lines based 
   on the equivalent width errors. 
   Basically, the values inferred by using the observed equivalent 
   widths better agree with those of the synthetic spectra. 
   In fact, the method based on the expected line strength 
   is not totally free from a bias that can heavily affect the 
   determination of microturbulent velocity. 
   Finally, I recommend to use 
   the classical approach (based on the observed equivalent widths) 
   to infer this parameter.
   In cases of full spectroscopical determination of all the atmospherical 
   parameters, the difference between the two approaches is reduced, leading 
   to an absolute  difference in the derived iron abundances of less than 0.1 dex.}
   {}


   \maketitle
%
%________________________________________________________________

\section{Introduction}

One of the major limitations in the model atmospheres based on 1-dimensional 
geometry is an unrealistic treatment of surface convection. 
Usually, the convective energy transfer is computed by adopting 
the classical mixing-length theory \citep{bohm59}, a parameterization 
aimed to by-pass the lack of a rigorous mathematical formulation 
for the convective  energy transfer. On the other hand, this parameterization cannot 
describe realistically the convective phenomena, ignoring their time-dependent
and non-local nature.\\ 
Basically, photospheric turbulent motions are divided in two typologies, 
according to the size of the turbulent element: if this is smaller 
than the mean free path of the photons we call it {\sl microturbulence}, 
otherwise {\sl macroturbulence}.\\ 
Such motions broaden the spectral lines, and the necessity to parametrize 
these effects arises from the discrepancy between 
empirical and theoretical Curves of Growth (COG) on their flat part. 
Microturbulent velocity (hereafter $v_{t}$) is a free-parameter introduced to 
remove this discrepancy and to compensate for the missing broadening. Usually, $v_{t}$ 
is included during the spectral synthesis in the Doppler shift 
calculation, adding in quadrature $v_{t}$ and thermal Doppler line broadening.\\
Some general objections can be advanced on this simplification:\\ 
(1)~the distinction between micro and macroturbulence is only an arbitrary 
boundary, because based on the uncorrected assumption that the velocity fields 
work on a discrete scale, while these velocity fields rather run on a continuous scale;\\ 
(2)~the 1-D model atmospheres (usually adopted in the chemical analysis) 
include depth-independent microturbulent velocity. Observational hints point 
toward a different scenario, 
see e.g. the works about Arcturus by \cite{gray81} and \citet{takeda92};\\ 
(3)~the turbulent velocity distribution is generally modeled as isotropic and Gaussian; 
this is contradicted by the observation of the granulation on the solar 
surface; as pointed out by \cite{nordlund97} and \cite{asplund00}, the phenomenon 
of photospheric granulation is characterized by laminar (and not turbulent) 
velocities, due to strong density stratification;\\ 
(4)~mixing-length theory describes the general temperature structure of the 
photosphere, but it is unable to model some relevant convective effects, as the 
photospheric velocity field and the horizontal fluctuations in the 
temperature structure \citep{steffen02}.

The classical and empirical way to estimate $v_{t}$ is to erase any slope between 
the abundance A(Fe) 
\footnote{I adopted the canonical formalism A(Fe) = $\log_{10}(N(Fe)/N(H))$ + 12.}
and the reduced equivalent width $\log_{10}(EW/\lambda)$ (hereafter EWR, where EW is the 
equivalent width), because $v_{t}$ 
works preferentially on the moderated/strong lines, while the weak ones are more 
sensitive to abundance. Even if the least square fitting of these two quantities 
is usually used to infer $v_{t}$, this method requires that the two quantities 
have uncorrelated errors. This condition is not respected in this case, 
because  the errors in EWR and A(Fe) are obviously correlated, with the risk to
introduce a spurious slope in the A(Fe)--EWR plane.

To bypass this obstacle and provide a more robust diagnostic to infer $v_{t}$, 
\cite{magain84} proposed to derive $v_{t}$ by using the theoretical EW 
(hereafter EWT) instead 
of the observed one. 
In this paper, I investigated the impact of these two approaches to derive 
$v_{t}$, in order to evaluate their reliability and to estimate the
best one to use in the typical chemical abundance analysis.

\section{Artificial spectra: general approach}
\label{arty}

The basic idea is to generate sets of synthetic spectra with a given and {\sl a priori} 
known $v_{t}$  (with inclusion of instrumental broadening and noise) and to analyze 
them with the two methods. Such a procedure 
allows to estimate the absolute goodness of the two methods (comparing the 
derived $v_{t}$ with the original one) but also the relative accuracy between 
the two approaches, in order to highlight systematic effects, if any. \\
Different sets of artificial spectra have been generated with the 
following method:\\ 
(1)~synthetic spectra for a giant star with $T_{\rm eff}$=4500 K, log~g=2.0, 
[M/H]=~--1.0 dex and $v_{t}$=2 km/s
have been computed with the spectro-synthesis 
code SYNTHE by R. L. Kurucz in its Linux version 
\citep[][]{sbordone04,sbordone05}, 
starting from LTE, plane-parallel ATLAS~9 model atmospheres and 
by adopting the line compilation by R. L. Kurucz 
\footnote{http://kurucz.harvard.edu/linelists.html}; 
the synthetic spectra cover a wavelength range between 5500 and 8000 
$\mathring{A}$;\\ 
(2)~the spectra have been convolved with a Gaussian profile in order to mimic 
the instrumental resolution of the modern echelle spectrographs. A spectral 
resolution of $\lambda/\delta\lambda$=~45000 has been adopted 
(analogous to that of several spectrographs as UVES, CRIRES, SARG);\\ 
(3)~the broadened spectra have been re-sampled to a constant wavelength step of 
$\delta$x=~0.02 $\mathring{A}$/pixel (similar to that of UVES echelle 
spectra);\\ 
(4)~poissonian noise has been added to the rebinned spectra, by using the 
IRAF task MKNOISE, in order to reproduce different noise conditions. 
SNR of 25, 40, 50, 70, 100 and 200 have been simulated. 
Due to the random nature of the noise, I resorted to MonteCarlo simulations, 
and for each value of SNR, 200 artificial spectra have been generated 
and analyzed independently.
\\ 
I selected from the Kurucz linelist $\sim$250
Fe~I transitions in the spectrum, taking into account only unblended 
lines or those with negligible blending, similarly to what is done on real 
observed spectra.
For all spectra, EWs have been measured with the automatic code 
DAOSPEC \citep{stetson08}.

\section{Determination of $v_t$ and caveats} 
\label{cav}

For all the synthetic spectra described above,
least square fits have been computed both in the 
A(Fe)--EWR and A(Fe)--EWT planes, choosing the value of $v_{t}$
that minimizes the slope. All the other parameters ($T_{\rm eff}$, log~g and 
[M/H]) were kept fixed, according to the original ones of the synthetic
spectra
\footnote{Note that this is a simplifying assumption. In fact, 
the most adopted method to perform a chemical analysis (the so-called 
{\sl full spectroscopic method}) is based on the following requirements: 
(i)~the iron abundance is independent of $\chi_{\rm ex}$ (constraint 
for $T_{\rm eff}$);
(ii)~the same abundance should be derived from single and ionized iron lines 
(constraint for log~g); 
(iii)~the iron abundance is independent of the line strength (constraint for 
$v_t$); 
(iv)~the model metallicity well reproduces the 
iron abundance (constraint for [M/H]).}. 

The most adopted proxy for the theoretical line strength 
can be written as $\log_{10}(gf)-\theta\chi_{\rm ex}$ 
\citep[see also Eq.~14.4 in][]{gray92}, 
where $\log_{10}(gf)$ is the oscillator strength, $\theta$ the Boltzmann factor 
(defined as $\theta$=~5040/$T_{\rm eff}$) and $\chi_{\rm ex}$ the excitation potential.
Note that EWT does not include terms strictly inferred by the observations but only 
related to the adopted model atmospheres ($\theta$) and to atomic data of the involved
transitions ($\log_{10}(gf)$ and $\chi_{\rm ex}$). 

\begin{figure}
   \centering
   \includegraphics[width=9cm]{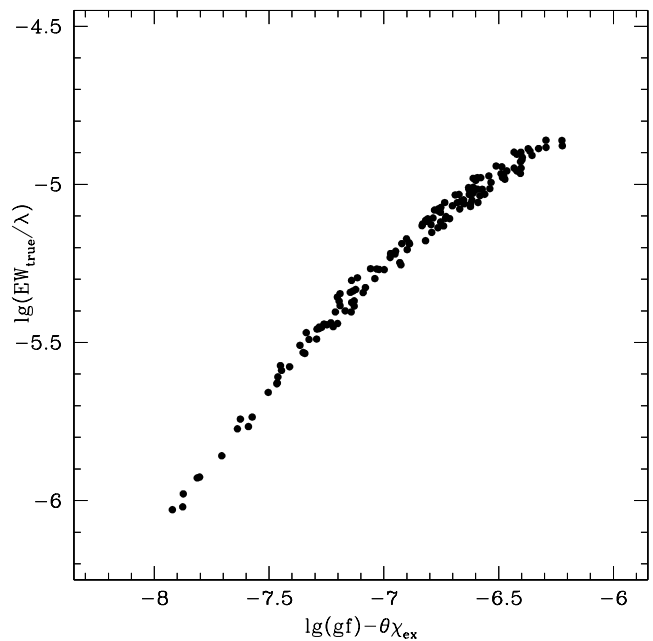}
      \caption{Behavior of the {\sl true} equivalent widths measured 
      in the synthetic spectrum (without  Poissonian noise) 
      computed with the atmospheric parameters of a giant star, as a function 
      of the quantity $\log_{10}(gf)-\theta\chi_{ex}$.}
         \label{the}
   \end{figure}

Fig.~\ref{the} shows the comparison between the true equivalent width computed 
for the synthetic spectrum of the giant star (before adding noise) and 
the proxy used for EWT. As appreciable in figure, the quantity 
$\log_{10}(gf)-\theta\chi_{\rm ex}$ is proportional 
to the true equivalent width of the synthetic spectrum 
(at least in the EW range shown in Fig.~\ref{the}), 
being a monotonic function of the true EW.
In this work, I adopt the quantity $\log_{10}(gf)-\theta\chi_{\rm ex}$ as the proxy 
for EWT because this is the formalism usually adopted in the chemical analysis 
that follow the method described by \citet{magain84}.

A threshold for the strongest lines has been applied:
only lines weaker than log(EW/$\lambda$)=~--4.85 have been included 
in the analysis, corresponding to $\sim$78 m$\mathring{A}$ at 5500 
$\mathring{A}$ and $\sim$113 m$\mathring{A}$ at 8000 $\mathring{A}$. 
Stronger lines (basically located along the flat/damped part of the COG) 
have been excluded in order to not introduce
additional effects (both observational and theoretical) that could 
affect the determination of $v_t$. In fact: (i)~strong 
features deviate from the Gaussian approximation (due to the development 
of Lorenztian wings), with the risk to under-estimate the EW 
\citep[as discussed in][see their Fig.~7]{stetson08}, and 
(ii)~these lines 
are affected by other effects, in addition to the microturbulent velocity, 
i.e. hyperfine/isotopic splitting, damping treatment 
and modeling of outermost atmospheric layers:
the inclusion of such lines could lead to a different $v_t$, 
very sensitive to the treatment of these effects for the iron lines, but it is 
not sure that the derived $v_t$ will be appropriate also for other elements
\citep[as discussed by][]{ryan98}. 
Note that, in principle, stronger lines can be taken into account in the abundance 
analysis (and for some elements like Ca or Ba only 
strong/damped transitions are often available), but they should 
be excluded from the determination of $v_t$.

It is noteworthy that the approach used in this work represents only a simple 
case with respect to the analysis of real spectra, because it is based on 
some simplifications:\\ 
(i)~large number of unblended lines. In real spectra, the number of lines 
is limited by the quality of the available oscillator strengths and by 
the wavelength coverage;\\ 
(ii)~only $v_{t}$ is adopted as a free parameter to determine, because the 
other atmospherical parameters are known a priori from the computation of the 
synthetic spectra;\\ 
(iii)~the used spectra are flat and their continuum (virtually) set to the 
unity. Hence, distortions of the continuum due to residual fringing, 
uncorrected merging of the echelle orders or residual of blaze function are 
not included;\\ 
(iv)~this analysis is limited to features that can be well reproduced by 
a Gaussian function, while in real cases some features with prominent 
damping wings can be included in the analysis by adopting a Voigt profile fitting 
(but taking into account the caveats discussed above for strong lines).

\section{Impact of SNR}
\label{resl}

\begin{figure}
   \centering
   \includegraphics[width=9cm]{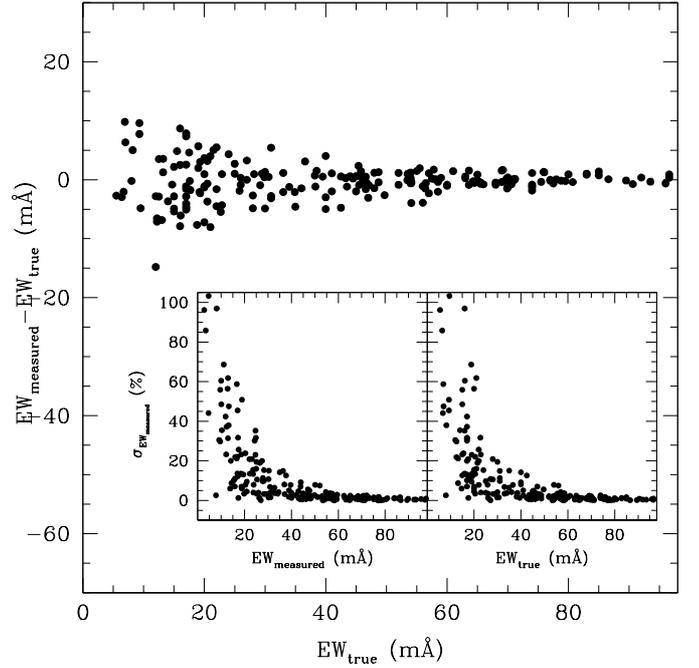}
      \caption{Difference between the measured and true EW in a 
      synthetic spectrum with SNR=25 as a function of the true EW. 
      In the left inset is shown the behavior of the error in 
      percentage of the measured EW as a function of the true EW, while 
      in the right inset is shown the same behavior as a function of the 
      measured EW.}
         \label{percent}
   \end{figure}

For each SNR, 200 artificial spectra were generated and analyzed independently, 
and the mean value of $v_{t}$ of each group of SNR was derived. 
An important step in this kind of analysis is a careful selection of the 
lines used to infer $v_t$, excluding features not well measured that could introduce 
a bias in the $v_{t}$ determination. In order to quantify the impact of inaccurate 
EWs in the $v_t$ estimate, I determined 
$v_{t}$ by considering different error thresholds and by taking into account only 
lines with errors $\sigma_{EW}$ less than 10, 20, 30 and 50\%; these errors have been 
computed by the DAOSPEC code from the standard deviation of the local flux residuals and 
represent a 68\% confidence interval on the derived value of the 
EW \citep[see][for details]{stetson08}. The discard of lines with inaccurate EWs 
have been performed individually for each artificial spectrum.
As sanity check, the EWs measured in a synthetic spectrum with SNR=25 have 
been compared with the true EWs of the original spectrum (before adding noise). 
The results are shown in Fig.~\ref{percent}; 
the inner panels show the 
behavior of the relative error in percentage as a function of the 
true and measured EW (left and right inset, respectively).
These plots confirm the reliability of the measured EWs and show that the 
errors in the EW measurement affect mainly the weak lines, while the strong 
($>$40 m$\mathring{A}$) lines are well measured with small uncertainties.

\begin{figure}
   \centering
   \includegraphics[width=9cm]{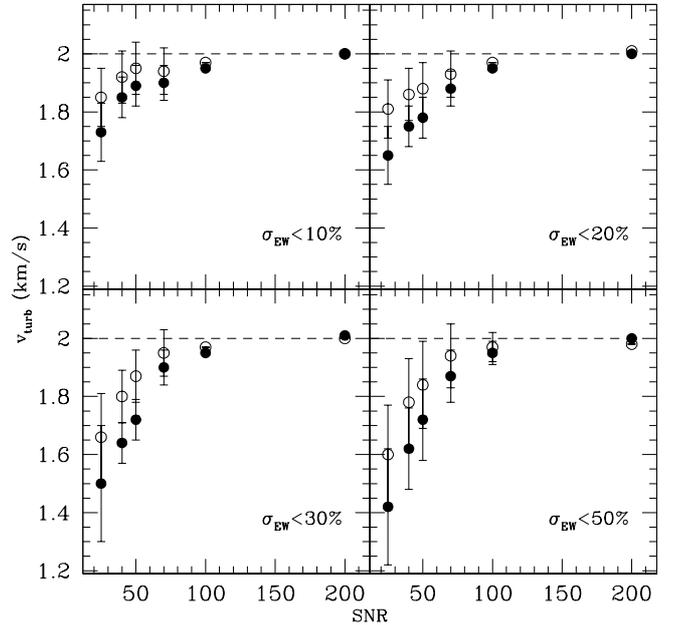}
      \caption{Behavior of the measured microturbulent velocity as a function 
      of SNR, derived with the classical method (empty circles) and 
      the prescription by \cite{magain84} (filled circles) for a 
      giant star. Each point is 
      the average value obtained by using 200 synthetic spectra of a given SNR and the 
      errorbars indicate the dispersion around the mean. Dashed lines indicate the 
      original value of $v_{t}$ of the synthetic spectra. 
      Each panel shows the results obtained by considering different thresholds 
      for the residual errors in the EW measurement.
              }
         \label{vt1}
   \end{figure}

The results concerning the determination of $v_t$ with the two methods for the 
sample of synthetic spectra described above are plotted in Fig.\ref{vt1}, 
where empty and filled circles indicate the values of $v_{t}$ derived by using 
the EWR method and EWT one, respectively. The errorbars indicate the 
dispersion around the mean of each set of simulations of a given SNR.
For sake of clarity, the original value of $v_{t}$ of the synthetic spectra 
is plotted as a dashed line. 
Typically, the dispersion around the mean in each group 
ranges from a few hundredths of km/s (for SNR=~100 and 200)
up to 0.2 km/s for SNR=~25 and considering lines with residual errors less than 50\%; 
note that such dispersions are basically compatible with the corresponding 
errors of the computed slopes
\footnote{For a sub-sample of synthetic spectra, the analysis has been 
repeated by using the true EWs instead of EWT as defined previously. The results 
are very consistent with those obtained by using EWT, in agreement with the 
discussion of Sect.~\ref{cav} and the finding of Fig.~\ref{the}.}.

In case of relatively high ($\ge$70) SNR, the two methods provide very similar 
results and agree with the $v_{t}$ of the original synthetic spectrum. In fact, 
increasing the SNR, the errors in the EW measurements are reduced and 
the discrepancy between the two approaches is erased. On the other hand, spectra 
with SNR$<$70 show $v_{t}$ always lower than the original value and the 
discrepancy increases when also lines with high uncertainties are taken 
into account. Generally $v_{t}$ inferred by using EWT turns out to be 
systematically lower 
than that obtained with the classical method by 0.1 -- 0.2 km/s, approximately. 
The reasons of this systematic under-estimate will be explained 
in Sect.~\ref{bias}.\\ 
These findings highlight also the impact of measured EWs with high uncertainties: 
the inclusion of lines with EWs very uncertain (generally, the weakest lines 
in spectra with low SNR) worsens the discrepancy with the true $v_t$ of the 
original synthetic spectra. Thus, a careful selection of the used lines, 
considering the accuracy of their EWs, is crucial in the derivation of $v_t$, 
in order to exclude additional bias.\\ 
A simple analytical model can be derived by fitting the data shown in 
Fig.~\ref{vt1}. The absolute value $\delta_{v_t}$ of the difference between the 
derived $v_t$ and the true one can be expressed as 
$\delta_{v_t}$=~$f_{\sigma_{EW}}/SNR$, where $f_{\sigma_{EW}}$ is a 
linear function of the adopted error threshold $\sigma_{EW}$ 
(expressed in \%). 
The fit of the present dataset provides 
$\delta_{v_t}$=~$(0.14\cdot\sigma_{EW}+2.64)/SNR$ for $v_t$ derived 
by using EWR and $\delta_{v_t}$=~$(0.19\cdot\sigma_{EW}+5.16)/SNR$ 
when EWT is adopted.\\
I repeated these checks also with synthetic spectra computed with 
the typical parameters of a dwarf star ($T_{\rm eff}$=6000 K, 
log~g=4.0, [M/H]=~--1.0 dex and $v_{t}$=1 km/s), finding basically 
the same result.

\section{Is the EWT approach free from biases?}
\label{bias}

\begin{figure}
   \centering
   \includegraphics[width=9cm]{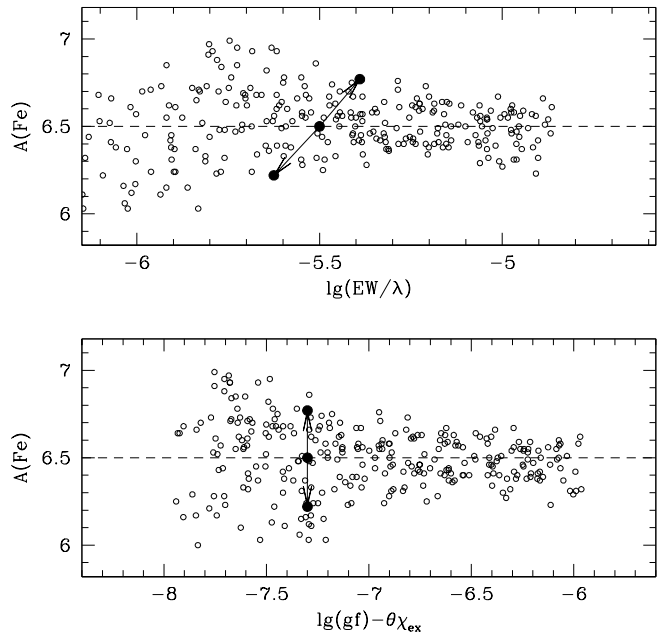}
      \caption{Behavior of the iron abundance as a function of  
      EWR (upper panel) and EWT (lower panel) for 
      a synthetic spectrum computed with $v_t$=~2 km/s. 
      Filled points show the shift of a given line in the two cases 
      due to the EW error.
              }
         \label{cay1}
   \end{figure}

An useful example to clarify the systematic offset between the two methods 
is to measure EWs of the red giant synthetic spectrum {\sl before adding noise}: 
all lines, regardless of their strengths, provide the same abundance. 
These EWs have been varied randomly by $\pm$1$\sigma$, 
where $\sigma$ was computed with the classical formula by \citet{cayrel88} 
($\sigma=1.5\cdot(SNR^{-1})\cdot\sqrt{FWHM\cdot\delta x}$, 
adopting SNR=~25, the wavelength step $\delta$x=~0.02 $\mathring{A}$/pixel and 
a FWHM of $\lambda$/R, corresponding to 0.12 $\mathring{A}$ at 5500 $\mathring{A}$
and 0.18 at 8000 $\mathring{A}$) and the analysis repeated as described above. 
The error is assumed the same regardless of the size of the EW.
Fig.~\ref{cay1} shows an example of this test, with the distribution of the Fe~I lines 
in the A(Fe)--EWR and A(Fe)--EWT planes
(upper and lower panel, respectively). 
Filled points show how a generic weak line (with initial A(Fe)=~6.5 dex) 
moves in the two diagrams because of the effect of EW errors. 
In the upper panel of Fig.~\ref{cay1} 
the effect due to the correlation between the error in EW measures 
and derived abundances is clearly visible, 
while the correlation is avoided by definition in A(Fe)--EWT plane: 
this is the main reason behind the use of EWT.

 \begin{figure}
   \centering
   \includegraphics[width=9cm]{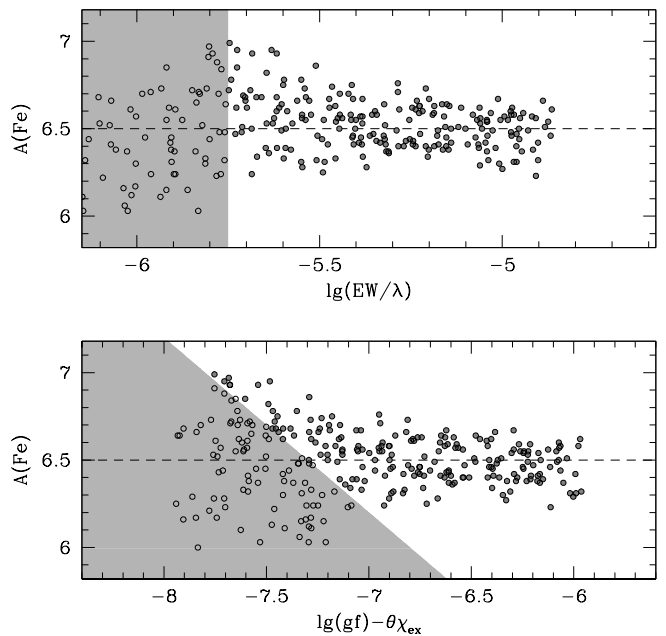}
      \caption{Same of Fig.~\ref{cay1} but highlighting the lines 
      detectable at SNR=~25 (grey points). Grey regions indicate 
      the location of the undetectable lines.
              }
         \label{cay2}
   \end{figure}

The mean slope, computed by taking
into account 200 MonteCarlo simulations of random error distribution, is 
of +0.13$\pm$0.02 in the A(Fe)--EWR plane, and --0.01$\pm$0.02
in the A(Fe)--EWT plane (ever keeping fixed the $v_t$ value at the original value 
of the synthetic spectrum, $v_t$=~2 km/s). Such slopes point out the need to increase $v_{t}$ 
(of at least 0.2 - 0.3 km/s) to erase the residual slope when $v_{t}$ is 
optimized by using the EWR. On the other side, adopting the EWT, no relevant 
change in $v_{t}$ are necessary. This finding reproduces exactly the bias 
discussed by \cite{magain84} (and shown in his Fig.~2)
but it seems to be contrary to the findings shown in Fig.~\ref{vt1}, where 
the two methods provide lower $v_{t}$ with respect to the true value.
In fact, \cite{magain84} demonstrated that $v_t$ is systematically
overestimated when derived with the classical method.\\
How can we explain such a discrepancy? In the real case (and also in the test 
performed with synthetic spectra, see Sect.~\ref{arty}), not all the lines can be measured and  
used to infer $v_{t}$. The effect of the {\sl slanting} movement of the lines 
shown in Fig.\ref{cay1} (upper panel) is critical  for very weak lines (where the 
errors are more relevant) but at low SNR not all these very weak features 
are detectable, because they are too noisy or indiscernible from the noise envelope. 
Hence, many weak lines that contribute to the positive slope in the 
A(Fe)--EWR plane will go out of the detectability threshold, reducing the 
positive slope.\\ 
For a given SNR, a reasonable and generally adopted threshold to measure spectral lines is
to assume 3 times the uncertainty provided by the Cayrel formula. Thus, 
for SNR=~25 we can measure absorption lines of at least 9 m$\mathring{A}$ 
and 11 m$\mathring{A}$ at 5500 and 8000 $\mathring{A}$, respectively. Because the 
difference is negligible along the entire coverage of the spectrum, I adopted 
conservatively a threshold of 11 m$\mathring{A}$.
Fig.~\ref{cay2} reports the same results of Fig.~\ref{cay1} but highlighting the 
lines really detectable assuming such a threshold (grey points), while the 
lines located in the grey shaded regions are excluded from the analysis.

The mean slope of the 200 MonteCarlo simulations, computed on the surviving 
lines, turns out to be of -0.10$\pm$0.02 in the 
A(Fe)--EWR plane, and -0.14$\pm$0.02 in the A(Fe)--EWT plane. 
When we take into account the effective possibility to detect very 
weak lines with large uncertainties, the situation turns in to the 
opposite of the case described in Fig.\ref{cay1}: 
\begin{itemize}
\item the threshold in the A(Fe)--EWT plane results to be diagonal, introducing 
a negative slope and the need to lower $v_{t}$ to erase any trend
\footnote{Note that the slope of the
boundary between detectable and undetectable lines 
is roughly orthogonal to
the shift of a line due to the correlation between the errors in EW and 
abundance.  Basically, the two slopes have opposite sign.}:\\
\item in both cases the derived $v_{t}$ will be lower than the original value 
of the synthetic spectrum (the same finding discussed above and shown in Fig.~\ref{vt1}).
\end{itemize}
 
In principle, all these effects are valid also for higher SNR, but become
negligible because the quoted uncertainties in the EW measurement are of 
only a few percent.

This experiment clarifies the origin of the systematic underestimate of $v_t$ 
with the two approaches (shown in Fig.~\ref{vt1}). When $v_t$ is derived from EWR, 
the correlation between the EWR and abundance errors moves the lines in an asymmetric 
way on the A(Fe)--EWR plane (upper panel of Fig.~\ref{cay1}), but the threshold 
of detectability of the lines will populate mainly the region with A(Fe)$>$6.5 dex 
with respect to other one (as appreciable in the upper panel of Fig.~\ref{cay2}), 
leading to a negative slope (and an underestimate of $v_t$).

When $v_t$ is derived from EWT, the threshold of the line selection shown in the 
lower panel of Fig.~\ref{cay2} increases the asymmetric distribution of the lines in 
the plane, ever leading a negative slope (as demonstrated by the average slopes in the 
two cases) and a slightly lower $v_t$ value.

Obviously, the inclusion of weak lines with inaccurate measures will worsen the 
discrepancy (lower panels in Fig.~\ref{vt1}).

Summarizing, $v_{t}$ inferred by using the A(Fe)--EWT plane are not totally 
free from {\sl observational} biases. In fact:\\ 
(a)~the use of A(Fe)--EWT plane partially erases the effect due to the 
correlation between errors in EW and abundances, 
but also introduces another bias due to the threshold of detectability 
of the lines (more relevant and crucial at 
low SNR), as described above;\\ 
(b)~errors in A(Fe) and EWT are still not totally independent from each other, because 
both quantities suffer in uncertainties from the adopted $\log_{10}(gf)$ values. 
In fact, an overestimate of the oscillator strength for a given line will 
provide an overestimate of EWT but also an underestimate of A(Fe), 
introducing a  negative, spurious slope when EWT is used to infer $v_{t}$. 
Note that this effect is secondary, if accurate laboratory oscillator strengths 
are taken into account. 
For instance, the typical errors of the Fe~I $\log_{10}(gf)$  in the 
critical compilation by \citet{fw06} are less than 25\%. 
Such errors translate in an iron abundance error of the same order of magnitude 
obtained with an error of EW of $\sim$25\% or less.  
The correlation between the errors in $\log_{10}(gf)$ and 
abundances introduce a spurious slope in the plane A(Fe)-EWT, in a similar 
way to the spurious slope in the A(Fe)-EWR plane due to the 
correlation between abundance and EWR discussed above. 
Otherwise, one needs to consider also this secondary bias only if inaccurate/inhomogeneous 
atomic data for iron lines are included in the linelist;\\
(c)~in the case where the temperature has been inferred by imposing the 
excitational equilibrium (erasing any slope in A(Fe)--$\chi_{\rm ex}$ plane), also 
$\theta$ will be affected (in a stronger way for low SNR spectra) by the 
errors in EW and therefore also EWT will be affected by the same bias, by definition. \\ 
Note that these last two aspects have not been deeply 
investigated in the present work

\section{Impact of the number of lines}

Previous results are based on a large number of Fe~I lines (250), 
allowing to well sample both weak and moderately strong features. Such an 
assumption allows to avoid spurious slopes (regardless to the adopted method) 
due to inadequate sampling of line strengths.\\
Some mid/high-resolution multi-object spectrographs offer wavelength 
coverages that limit the number of available Fe~I lines.  

A simple test has been performed in order to clarify how the number of lines is critical. 
50 Fe~I lines have been extracted randomly from the measured lines in a red giant
synthetic spectrum with noise SNR=25 and $v_{t}$ was redetermined with the two 
methods. Such a procedure has been repeated by using 200 MonteCarlo extractions. 
The average values of the $v_{t}$ distributions is very similar to those 
obtained above but with a larger dispersion ($\sigma\sim$0.3--0.4 km/s) 
with respect to that obtained in the case described in Section.~\ref{resl}.\\ 
If the same check is performed with a synthetic spectrum with SNR=100, 
the final distributions share dispersions around the mean of about 
$\sigma\sim$0.15--0.2 km/s, in both cases, while in the case discussed in Section~\ref{resl} 
the dispersion was very small (see Fig.~\ref{vt1}). 
Also in this case, such dispersions are larger than those inferred from EWR and EWT 
when a large number of transitions is used.

The most discrepant $v_t$ values in the derived distributions are obtained when 
an inadequate sampling of the weak and strong lines is adopted. In fact, if the 
experiment is repeated imposing the condition that the 50 random lines are extracted 
by sampling equally the entire EW range (thus providing a statistically significant sample 
of weak and strong lines), the dispersion around the mean of the derived distribution of 
$v_t$ is reduced by about a factor of 1.5 and 2 for SNR=25 and 100 respectively. 
In this case, the distributions show dispersions around the mean that
are barely consistent with those obtained with the entire linelist. 
I can conclude that  
the number of used lines and the relative sampling between 
weak and strong lines are crucial elements that can affect the 
determination of $v_{t}$ more severely than the choice of the EWR or EWT method 
in the analysis.

\section{Impact of the continuum placement}

As already pointed out above, the test performed with synthetic spectra 
neglects uncertainties arising from the continuum location, because 
the continuum in artificial spectra is set at unity and totally flat, 
while in real spectra several additional effects affect its correct 
identification. In order to clarify how much 
this source of uncertainty in the EW measurements impacts $v_{t}$, 
I performed two simple tests by considering synthetic spectra with SNR=~25:\\ 
(1)~the EWs have been remeasured by adopting a higher and lower continuum 
level, with respect to that derived in the previous experiment. The 
variations in the continuum level are of $\pm$5 and $\pm$10\%, translating 
to a systematic (positive or negative depending by the sign of continuum 
variation) offset of the derived EWs (the same finding has been discussed 
in \citet{stetson08}, see their Fig. 2). 
The differences between $v_t$ derived with the EWR and EWT methods 
are similar to those obtained with the original continuum level;\\ 
(2)~the synthetic spectra  have been multiplied with a sinusoidal curve, 
in order to mimic a fringing-like pattern, similar to that observed in real 
spectrum. Also in this case, no relevant difference (within the uncertainties) 
with the original analysis is found.

The continuum tracement (at least in case where the continuum is defined globally 
along the entire spectrum) does not affect dramatically the results described in 
Sect.~\ref{resl} 
about the relative difference between $v_t$ derived with EWR and EWT methods.

\section{A real case: UVES spectra of LMC giants}

In order to show a case based on real spectra, I derived $v_{t}$ with the 
two methods for some Large Magellanic Cloud giants stars observed with UVES
\citep{mucciarelli09}, with typical SNR=~50 and originally analyzed following 
the prescriptions by \cite{magain84}. 

Microturbulent velocities computed with the classical method 
(and fixing the other parameters) are  higher by $\sim$0.2 -- 0.25 km/s 
than those obtained with the prescription
by \cite{magain84}; this difference translates to an iron abundance 
$\sim$0.15 dex lower when the classical approach is adopted.

A point to recall here is that the $T_{\rm eff}$ and $v_{t}$ can be correlated 
if these quantities are derived with the classical full spectroscopic method. 
In fact, the majority of weak lines have high $\chi_{\rm ex}$, while several 
strong transitions have low excitation potentials, leading to 
a correlation between $\chi_{\rm ex}$ and the strength of the lines. 
Also, variations in $T_{\rm eff}$ and $v_{t}$ change differently 
single and ionized iron lines abundances, needing to re-adjust the 
surface gravity.
The higher $v_{t}$ computed with the classical method implies the need  
to adjust spectroscopically both $T_{\rm eff}$ and log~g, by increasing them and 
hence reducing the iron abundance difference (to less than 0.1 dex).

Summarizing, {\sl the absolute difference in iron abundance computed 
with the two approaches is reduced (less than 0.1 dex) 
if a full spectroscopic determination of all atmospherical parameters is performed}. 
Therefore, if one or more parameters are fixed to a certain value, this 
{\sl compensation} does not take place and the absolute difference can be higher.

\section{Determination of $v_t$: the best strategy}

I propose an easy strategy to minimize the effect of the bias discussed 
above, 
deriving the $v_t$ from EWR, being the approach that gives 
the better results with respect to the true $v_t$ of the synthetic 
spectra used in Sect.~\ref{resl}:

(i)~being the determination of $v_t$ strongly affected by the 
correlated errors in abundances and EWs, the lines with 
high uncertainties in the EW measurement need to be discarded.
In fact, as demonstrated in Fig.~\ref{vt1}, the inclusion of lines with 
inaccurate EWs leads to highly uncorrected $v_t$ with respect to the true value. 
Basically, the use of lines with EW errors less than 10\% provides 
a good agreement with the $v_t$ value of the synthetic spectra. 
The reader can use the relations provided in Sect.~\ref{resl} to 
estimate the amplitude of the bias in $v_t$ as a function of the 
adopted $\sigma_{EW}$ cutoff. 
Note that the computation of the EW uncertainty is not provided by 
all the codes developed to measure EWs, but a standard error about the goodness 
of the fit for each measured
line is recommended, through the standard deviation of the local 
flux residuals after the Gaussian fitting
(as performed by DAOSPEC) or by adopting other formulae, i.e. the Cayrel formula 
or that proposed by \citet{ramirez};

(ii)~as shown in Fig.~\ref{vt1}, even if an adequate lines selection 
is applied, a residual discrepancy between the derived and true $v_t$ 
of $\sim$0.15 km/s remains in the most critical case of SNR=~25. 
A possible method to reduce this residual discrepancy is to 
perform a linear fit obtained taking into account the uncertainties 
in both EWs and abundances (following the approach by \citet{press} 
as implemented in the FORTRAN routine FITEXY).
Errors in EWs have been estimated as described in Sect.~\ref{resl}, 
while the error in the abundance of each individual line has been 
derived by computing the abundance with the EW varied of $\pm$1$\sigma_{EW}$. 
This method to perform the linear fit allows to obtain values of $v_t$
consistent within the errors with the true value of the synthetic spectra 
(see the first panel of Fig.~\ref{vt1}), reducing the residual discrepancy.
Usually in the chemical analysis, the slope in A(Fe)---EWR plane is computed 
neglecting the uncertainties in the two variables 
(because this procedure needs the computation of the error in both EW and 
abundance for each measured line)
but this approach 
\citep[or other approaches to find the linear relationship in data with uncertainties in both directions, see e.g.][]{hogg}
is recommended. 

\section{Conclusions}
I draw the following results from the above test:\\ 
(1)~the method to infer the microturbulent velocity from the EWT 
provides the same results (within the quoted uncertainties) of 
the classical method in case of moderate/high SNR ($\ge$70);\\ 
(2)~in case of SNR$<$70 both methods underestimate the microturbulent 
velocity and when low (25) SNR spectra are analyzed a better agreement 
with the true $v_{t}$ of the synthetic spectra is found with the 
classical method based on EWR;\\ 
(3) the discrepancy with the true $v_t$ value (regardless of the 
adopted method) worsens when EWs with high uncertainties (generally, the 
weakest lines in low SNR spectra) are taken into account. 
This finding points out the need to estimate the quality of the measured 
EWs for each individual line (a procedure usually not performed in the 
chemical analysis), in order to reduce this systematic underestimate;\\ 
(4)~these results are valid for both giant and dwarf stars, under the 
assumptions that the strong lines that deviate significantly from the 
Gaussian approximation are excluded from the analysis;\\ 
(5)~the number of used lines  and the adequate sampling of weak and strong lines 
in the case of a small number of available iron lines affect the determination of $v_{t}$ 
more than the adopted approach to infer this parameter;\\ 
(6)~the spectroscopic optimization of all the atmospheric parameters 
simultaneously is recommended, in order to compensate errors in 
the determinations of the parameters.

\acknowledgements
I thank the referee (Matthias Steffen) whose comments and suggestions 
have improved significantly the paper.
I warmly thank Elena Pancino for her useful suggestions and 
a careful reading of the manuscript. 

%_____________   \begin{figure*}

% Online Material
%_____________________________________________________________
%        Online appendices have to be placed at the end, after
%                                        \end{thebibliography}
%-------------------------------------------------------------


\begin{thebibliography}{}
   
  \bibitem[Asplund et al.(2000)]{asplund00}
  Asplund, M., Nordlund, A., Trampedach, R., Allende Prieto, C., 
  \& Stein, R. F., 2000, A\&A, 359, 729
  \bibitem[B\"{o}hm-Vitense(1958)]{bohm59} 
      B\"{o}hm-Vitense, E., 1958, ZA, 46, 108
      \bibitem[Cayrel(1988)]{cayrel88}
      Cayrel, R., 1988, in IAU Symposium, Vol.132, The Impact of very high 
      S/N SPectroscopy on Stellar Physics, ed. G. Cayrel de Strobel \& 
      M. Spite, 345
      \bibitem[Fuhr \& Wiese(2006)]{fw06}
      Fuhr, J. R., \& Wiese, W. L., 2006, JPCRD, 35, 1669
  \bibitem[Gray(1981)]{gray81} 
   Gray, D. F., 1981, ApJ, 245, 992
  \bibitem[Gray(1992)]{gray92} 
    Gray, D. F., 1992, The Observation and analysis of stellar spectra, 
    Cambridge Astrophys. Ser.
   \bibitem[Hogg, Bovy \& Lang(2010)]{hogg}
   Hogg, D. W., Bovy, J., \& Lang, D., 2010, arXiv1008.4686
  \bibitem[Magain(1984)]{magain84}
  Magain, P., 1984, A\&A, 134, 189
  \bibitem[Mucciarelli et al.(2009)]{mucciarelli09}
  Mucciarelli, A., Origlia, L., Ferraro, F. R., \& Pancino, E., 2009, 
  ApJ, 659, 134L
  \bibitem[Nordlund et al.(1997)]{nordlund97}
  Nordlund, A., Spruit, H. C., Ludwig, H.-G., \& Trampedach, R., 1997, 
  A\&A, 328, 229
  \bibitem[Press et al.(1992)]{press}
  Press, W. H.,Teukolsky, A. A., Vetterling, W. T., \& Flannery, B. P., 
  Numerical Recipes, 2nd edn. Cambridge Univ. Press, Cambridge
  \bibitem[Ramirez et al.(2001)]{ramirez}
  Ramirez, S. V., Cohen, J. G., Buss, J., \& Briley, M. M., 2001, 
  AJ, 122, 1429
  \bibitem[Ryan(1998)]{ryan98}
  Ryan, S., A\&A, 331, 1051
  \bibitem[Sbordone et al.(2004)]{sbordone04}
   Sbordone, L., Bonifacio, P., Castelli, F., \& Kurucz, R. L. 2004, Memorie
   della Societa Astronomica Italiana Supplement, 5, 93
  \bibitem[Sbordone et al.(2005)]{sbordone05}
  Sbordone, L., 2005, Memorie della Societa Astronomica Italiana Supplement, 8, 61
  \bibitem[Steffen \& Holweger(2002)]{steffen02}
  Steffen, M., \& Holweger, H., 2002, A\&A, 387, 258
  \bibitem[Stetson \& Pancino(2008)]{stetson08}
  Stetson, P. B., \& Pancino, E., 2008, PASP, 120, 1332
  \bibitem[Takeda(1992)]{takeda92}
  Takeda, Y., 1992, A\&A, 253, 487
  

\end{thebibliography}
\end{document}